\documentclass[aps,prl,reprint,showpacs,longbibliography]{revtex4-1} 
\usepackage{graphicx}  % needed for figures
\usepackage{dcolumn}   % needed for some tables
\usepackage{amssymb}
\usepackage{amsmath}
\usepackage{physics}
\usepackage{natbib}
\usepackage{xcolor}
\usepackage{soul}
\usepackage{siunitx}
\usepackage{float}
\usepackage{esdiff}
\usepackage{bbold}

\colorlet{RED}{red}

\begin{document}
%Title of paper

\title{Generating two continuous entangled microwave beams using a dc-biased Josephson junction}

\author{A. \surname{Peugeot}$^{1}$}

\author{G. \surname{M\'enard}$^{1}$}

\author{S. \surname{Dambach}$^{2}$}

\author{M. \surname{Westig}$^{1}$}

\author{B. \surname{Kubala}$^{2,3}$}

\author{Y. \surname{Mukharsky}$^{1}$}

\author{C. \surname{Altimiras}$^{1}$}

\author{P. \surname{Joyez}$^{1}$}

\author{D. \surname{Vion}$^{1}$}

\author{P. \surname{Roche}$^{1}$}

\author{D. \surname{Esteve}$^{1}$}

\author{P. \surname{Milman}$^{4}$}

\author{J. \surname{Leppäkangas}$^{5}$}

\author{G. \surname{Johansson}$^{6}$}

\author{M. \surname{Hofheinz}$^{7}$}

\author{J. \surname{Ankerhold}$^{2}$}
\email{email: joachim.ankerhold@uni-ulm.de}

\author{F. \surname{Portier}$^{1}$}
\email{email: fabien.portier@cea.fr}

\affiliation{$^{1}$ DSM/IRAMIS/SPEC, CNRS UMR 3680, CEA, Universit\'e Paris-Saclay, 91190 Gif sur Yvette, France}
\affiliation{$^{2}$ Institute for Complex Quantum Systems and IQST, University of Ulm, 89069 Ulm, Germany}
\affiliation{$^3$ Institute of Quantum Technologies, German Aerospace Center (DLR), 89069 Ulm, Germany} 
\affiliation{$^{4}$ Université de Paris, Laboratoire Matériaux et Phénomènes Quantiques, CNRS, 75013 Paris, France}
\affiliation{$^{5}$ Institute of Physics, Karlsruhe Institute of Technology, 76131 Karlsruhe, Germany}
\affiliation{$^{6}$ Department of Microtechnology and Nanoscience (MC2), Chalmers University of Technology, SE-412 96 Gothenburg, Sweden}
\affiliation{$^{7}$ Institut  Quantique, Université de Sherbrooke, Sherbrooke, Canada}

\begin{abstract}
We show experimentally that a dc-biased Josephson junction in series with two microwave resonators emits entangled beams of microwaves leaking out of the resonators. In the absence of a stationary phase reference for characterizing the entanglement of the outgoing beams, we measure second-order coherence functions for proving entanglement up to an emission rate of 2.5 billion photon pairs per second. The experimental results are found in quantitative agreement with theory, proving that the low frequency noise of the dc bias is the main limitation for the coherence time of the entangled beams. This agreement allows us to evaluate the entropy of entanglement of the resonators, and to identify the improvements that could bring this device closer to a useful bright source of entangled microwaves for quantum-technological applications.

\end{abstract}

\date{\today}

\maketitle
%\section{Introduction}
Although the link between electrical transport and emission of radiation has been understood since the invention of electrical lamps, its complete description in the context of quantum  conductors requires a comprehensive treatment of the conductor itself, of the charge reservoirs connected to it, and of the electromagnetic modes of the environment that sustain radiation. Despite numerous achievements  \cite{cottet2015, dmytruk2016, mora2017, altimiras2016, grimsmo2016,leppakangas2014, leppakangas2013}, a full understanding is still missing in the general case. A Josephson junction, connected to a small number of modes originally in the vacuum state, provides a simple model system for this physics. For a dc bias $V$ smaller than the gap voltage $2\Delta/\mathrm{e}$, no quasiparticle excitation can absorb the energy $2\mathrm{e}V$ yielded by the biasing circuit upon the tunneling of a Cooper pair. As a result, a dc current flows through the junction only if this energy can be absorbed by creating photons in the environmental modes \cite{averin1990,ingold1992,holst1994,hofheinz2011}. Consequently, the properties of the emitted light depend both on the control voltage and on the coupling of the junction to the modes, described by their impedance $\mathrm{Re}[Z(\nu)]$. Previous experiments have shown that shaping $\mathrm{Re}[Z(\nu)]$ by microwave engineering allows for the creation of various non-classical states of light \cite{grimm2019,westig2017,rolland2019} and have thereby led to the emergence of the field
of \textit{Josephson photonics}. In the case where the junction is connected to two-modes at frequencies $\nu_\mathrm{a}\neq\nu_\mathrm{b}$, setting the voltage bias such that $2\mathrm{e}V=h\nu_\mathrm{a}+h\nu_\mathrm{b}$ results in the emission of photon pairs, with one photon created in each mode for each Cooper pair tunneling. The experimental observation of this pair-emission mechanism demonstrated that the beams leaking out of the two resonators have non-classical population correlations \cite{westig2017} but did not provide information on their quantum phase correlations. Are the two microwave beams entangled as one could expect ? If so, what is the meaning of two-mode entanglement in absence of any  phase reference ? In order to characterize the precise nature of the quantum correlations present in this unique non-classical two-beam source, we have built a new measurement setup able to probe entanglement between the output microwave beams.

The two resonator fields are coherently driven by the Josephson junction. This driving is described by an effective two-mode squeezing Hamiltonian, resulting in non-local quadrature correlations of the emitted light which we evidence experimentally. However, in our experiment, the absence of a well-defined phase reference associated to the dc bias and the thermal diffusion of the squeezing angle \cite{leppakangas2014} forbids the use of standard techniques to reveal these correlations \cite{debuisschert1989, heidmann1987,brida2009,forgues2014,flurin2012,flurin2014,PhysRevLett.113.110502,PhysRevLett.124.140503}. To demonstrate entanglement in a such scenario, without phase stability, we 
rely on the measurement of second-order correlation functions following a detection scheme inspired by Franson interferometry~\cite{franson1989}. A simple entanglement witness based on these correlators allows us to prove the entanglement between the outgoing fields.

\section{Circuit model and photon correlations}

\begin{figure}[hb]
\begin{center}
\includegraphics[width = 8.6 cm]{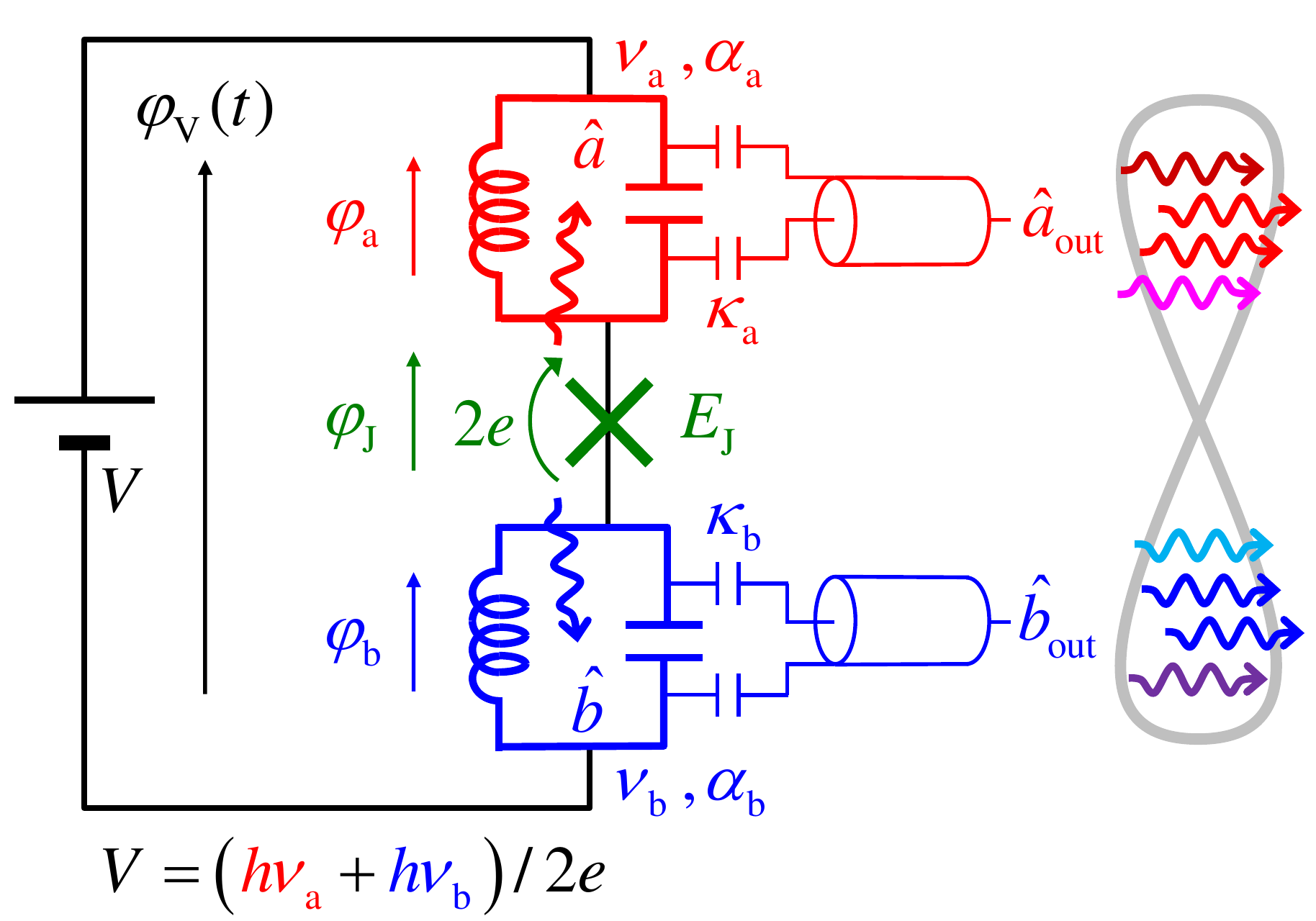}    
\caption{\textbf{Principle of the experiment}: A simple superconducting circuit made of a Josephson junction (green cross - Josephson energy $E_\mathrm{J}$) in series with two microwave resonators (red and blue LC circuits) having different frequencies $\nu_\mathrm{a,b}$, similar  coupling $\alpha_\mathrm{a,b}$ to the junction (see text), and similar energy leak rates $\kappa_\mathrm{a,b}$, is dc-biased by a voltage source V. Phases $\varphi$ entering the circuit Hamiltonian are also indicated. At a particular voltage (indicated at the bottom), each Cooper pair $2\mathrm{e}$ passing through the junction (green arrow) creates in a coherent way a pair of photons (vertical red and blue wiggles) in the two resonators; entangled microwave fields $a$ and $b$ grow in the resonators, as well as two stationary entangled microwave beams $a_\mathrm{out}$ and $b_\mathrm{out}$. By measuring correlations between $a_\mathrm{out}(t)$ and $b_\mathrm{out}(t)$, we experimentally demonstrate this effect.}
\label{fig1}
\end{center}
\end{figure}

The model circuit that we implement in this paper is presented in Fig.~\ref{fig1}: The series combination of a Josephson junction with Josephson energy $E_\mathrm{J}$, and of two resonators with different frequencies $\nu_\mathrm{a,b}$ but similar characteristic impedances $Z_\mathrm{a,b}$ and energy leak rates $\kappa_\mathrm{a,b}=2\pi \nu_\mathrm{a,b}/Q_\mathrm{a,b}$, is biased by an ideal dc voltage source $V$, 
\begin{equation}
 \varphi_J + \varphi_\mathrm{a}+ \varphi_\mathrm{b} =\varphi_{V}\equiv  2\mathrm{e}Vt/\hbar.
\end{equation}

Up to the zero-point energy of the resonators, the resulting time-dependent Hamiltonian of the circuit reads
\begin{equation}
\hat{H} = h\nu_\mathrm{a} \hat{n}_\mathrm{a} + h\nu_\mathrm{b} \hat{n}_\mathrm{b} -E_\mathrm{J} \cos( \hat{\varphi}_\mathrm{a} + \hat{\varphi}_\mathrm{b}-2\mathrm{e}Vt/\hbar).
\label{initial Hamiltonian}
\end{equation}
Here $\hat{\xi}\in \{\mathrm{a,b}\}$ is the annihilation operator of mode $\xi$, $\hat{n}_{\xi}=\hat{\xi}^\dagger \hat{\xi}$ is the corresponding number operator, and $\hat \varphi_{\xi} = \sqrt{\alpha_{\xi}}(e^{2i \pi \nu_{\xi} t} \hat{\xi}^\dagger +\mathrm{h.c.})$ with $\alpha_{\xi}=\pi Z_{\xi}(2\mathrm{e})^2/h$.
The coupling constant $\alpha_\xi$ for the interaction between the junction and mode $\xi$ plays the same role as the fine structure constant in quantum electrodynamics.
The $\cos(\varphi)$ term of Eq. \ref{initial Hamiltonian} may be expanded exactly, yielding an infinite series of terms oscillating at $2\mathrm{e}V/h+m\nu_\mathrm{a}+n\nu_\mathrm{b}$ with $\{m,n\}\in\mathbb{Z}^2$. The particular case of interest in this work is the resonance condition $2\mathrm{e}V\simeq h\nu_\mathrm{a}+h\nu_\mathrm{b}$, for which the energy delivered by the voltage source for each tunneling Cooper pair is entirely converted into a pair of photons, one in each mode. Moving to a frame rotating with the frequency $\nu_\mathrm{a}+\nu_\mathrm{b}=2\mathrm{e}V/h-\delta$ and performing a rotating-wave approximation, for a detuning $\delta$ small enough compared to $\nu_\mathrm{a},\nu_\mathrm{b}$, we arrive  at the effective Hamiltonian
\begin{equation}
\hat H_\text{RWA} \!=\! \frac{B}{2}:\!\sum_{k,l}^\infty\frac{\!(-1)^{k+l} (\alpha_\mathrm{a} \hat n_\mathrm{a}) ^k (\alpha_\mathrm{b} \hat n_\mathrm{b})^l\!}{k!l!(k+1)!(l+1)!} e^{-2i\pi\delta t}\hat a^\dagger \hat b^\dagger\!:+\text{h.c}
\label{HRWA}
\end{equation}
in terms of the creation and annihilation operators, or equivalently at
\begin{equation}
\hat H_\text{RWA} =\frac{B}{2}\left( e^{-2i\pi\delta t}  \hat{a}^\dagger \hat{b}^\dagger \hat{L}_\mathrm{a} \hat{L}_\mathrm{b} +\text{h.c} \right),
\label{HRWA2}
\end{equation}
which is more convenient for numerical evaluation. Here, $B = E_{\mathrm{J}}^{\ast}  \sqrt{\alpha_\mathrm{a} \alpha_\mathrm{b}}$ with $E_{\mathrm{J}}^{\ast} = E_{\mathrm{J}} e^{- \frac{\alpha_\mathrm{a} + \alpha_\mathrm{b}}{2}}$ is a reduced Josephson energy {\cite{schon1990,grabert1998,joyez2013,gramich2013,dambach2015,martinez2019,rolland2019}}, the colons indicate normal ordering, and $\hat{L}_{\xi} = \sum_{n_{\xi} = 0}^{\infty} L_{n_{\xi}}^{(1)} (\alpha_{\xi}) / (n_{\xi} + 1) |n_{\xi} \rangle \langle
n_{\xi}|$ is a diagonal operator involving the generalized Laguerre polynomials $L_n^{(1)}$. In the regime of weak coupling of the modes $\alpha_{\xi} \ll 1$, and of low resonator populations $\alpha_{\xi} \langle \hat{n}_{\xi} \rangle \ll 1$, the $(k, l) \neq (0, 0)$ terms may be neglected in Eq. \ref{HRWA} and $\hat{L}_{\xi} \simeq \hat{\mathbb{1}}$ in Eq. \ref{HRWA2}, such that $\hat{H}_\text{RWA}$ reduces to the two-mode squeezing Hamiltonian

\begin{equation}
\hat H_\text{TMS} = \frac{B}{2}e^{-2i\pi\delta t}\hat a^\dagger\hat b^\dagger+\text{h.c}.
\label{HTMS}
\end{equation}

The coherent driving term $\hat a^\dagger \hat b^\dagger$ with absolute amplitude $B$ creates photon pairs in $a$ and $b$ and entangles the two resonators, as we will show. The corresponding fields leak out at rates $\kappa_\mathrm{a,b}$, which brings the resonators in a stationary state and form two entangled beams of light centered on frequencies $\simeq \nu_\mathrm{a}+\delta/2$ and $\simeq \nu_\mathrm{b}+\delta/2$. A natural dimensionless driving strength is thus $\beta=B/\hbar\sqrt{\kappa_\mathrm{a} \kappa_\mathrm{b}}$. An input-output model \cite{leppakangas2014} allows us to relate the properties of the outgoing fields $\hat a_\text{out}(\nu),\hat b_\text{out}(\nu')$ to $\beta$, $\kappa_\mathrm{a,b}$, and $\delta$. The field quadratures at frequency $\nu$ in the $a$ band are found to be correlated with field quadratures at frequency $\nu'=2\mathrm{e}V/h-\nu$ in the $b$ band: A sum of properly chosen quadratures of $\hat a_\text{out}(\nu)$ and $\hat b_\text{out}(\nu')$ displays fluctuations below their value in the vacuum state. This two-mode-squeezing is predicted to be the largest at the maximum of the emission spectral density.

Detection of such a squeezing is usually achieved by filtering the $a_\text{out}(\nu)$ and $b_\text{out}(\nu')$ fields in narrow frequency bands and correlating their quadratures after demodulation  \cite{debuisschert1989, heidmann1987,brida2009,forgues2014,flurin2012,flurin2014,PhysRevLett.113.110502,PhysRevLett.124.140503}. In the ideal noiseless bias case, a non-zero expectation value of the first-order correlation function (second-order in the fields) $\langle \hat a_\text{out}\hat b_\text{out}(t)\rangle\propto e^{-2i\pi\delta t}$ (where $\hat \xi_\text{out}(t)=\int\hat \xi_\text{out}(\nu)e^{2i\pi(\nu-\nu_{\xi}) t}d\nu$) reveals the correlations between $\hat a_\text{out}(\nu)$ and $\hat b_\text{out}(\nu')$. However, in our experiment we cannot implement this protocol because of the thermal fluctuations of the voltage bias, and consequently of $\delta$, which blur correlations between $\nu$ and $\nu'$ faster than we can measure them.  Instead, we resort to a measurement scheme analog to the one originally developed by Franson~\cite{franson1989}. We use the particular second-order correlation function (fourth order in the fields) $\langle \hat a_\text{out}^\dagger(t)\hat b_\text{out}^\dagger(t) \hat a_\text{out}(t+\tau)\hat b_\text{out}(t+\tau) \rangle$ in which the unknown phase of $ \hat a_\text{out}(t+\tau)\hat b_\text{out}(t+\tau)$ is compensated by the counter-rotating phase of $\hat a_\text{out}^\dagger(t)\hat b_\text{out}^\dagger(t)$ for time delays $\tau$ shorter than the frequency jitter auto-correlation time. This correlator, thus averages to a finite value for such time delays and makes it possible to define an entanglement witness for the propagating fields. This concept for the detection of entanglement is not limited to the present system and can generally be applied to any system lacking phase stability.

Starting from a theorem demonstrated in Ref.~\cite{wolk2014}, we developed an entanglement witness on the basis of the Franson-type correlator: for any separable state of $a_\text{out}$ and $b_\text{out}$ the following inequality holds
\begin{equation}
    \left|g^{(2)}_\phi(\tau)\right|\leq \sqrt{g_\mathrm{ab}^{(2)}(\tau)\times g_\mathrm{ab}^{(2)}(-\tau)}\equiv g_\mathrm{ab,sym}^{(2)}(\tau),
\label{inequality}
\end{equation}
where
\begin{equation}
    g^{(2)}_{\phi}(\tau) = \left<\hat{a}^\dagger_\mathrm{ out}(0)\hat{b}^\dagger_\mathrm{out}(0)\hat{a}_\mathrm{out}(\tau)\hat{b}_\mathrm{out}(\tau)\right>/D
    \label{g2phi}
\end{equation}
and
\begin{equation}
    g^{(2)}_\mathrm{ab}(\tau) = \left<\hat{a}^\dagger_\mathrm{out}(0)\hat{a}_\mathrm{out}(0)\hat{b}^\dagger_\mathrm{out}(\tau)\hat{b}_\mathrm{out}(\tau)\right>/D
    \label{g2ab}
\end{equation}
with $D=\kappa_\mathrm{a} n_\mathrm{a} \kappa_\mathrm{b} n_\mathrm{b}$ and $n_{\xi}=\langle \hat{n}_{\xi} \rangle$. Measuring a violation of inequality (\ref{inequality}) is thus a sufficient condition for $a_\text{out}$ and $b_\text{out}$ to be entangled. Both, $|g^{(2)}_\phi(\tau)|$ and $g_\mathrm{ab,sym}^{(2)}(\tau)$, initially decay quickly on a scale set by  the photon lifetime, which determines for how long photons created by the same tunneling event are observed. After this initial decay $g_\mathrm{ab,sym}^{(2)}(\tau) \rightarrow 1$ becomes constant, while the further decay of $|g^{(2)}_\phi(\tau)|$ reflects how phase correlations between $a$ and $b$ modes are scrambled by the low frequency noise. This implies that the witness can prove entanglement of the outgoing modes only if photons leave the resonator faster than phase coherence is washed out. As the dephasing time expected from a cold environment is in $\mu$s range, we design our resonator with energy decay times of a few nanoseconds.

\section{Experimental setup}

\begin{figure}[ht]
\begin{center}
\includegraphics[width =  8.6 cm]{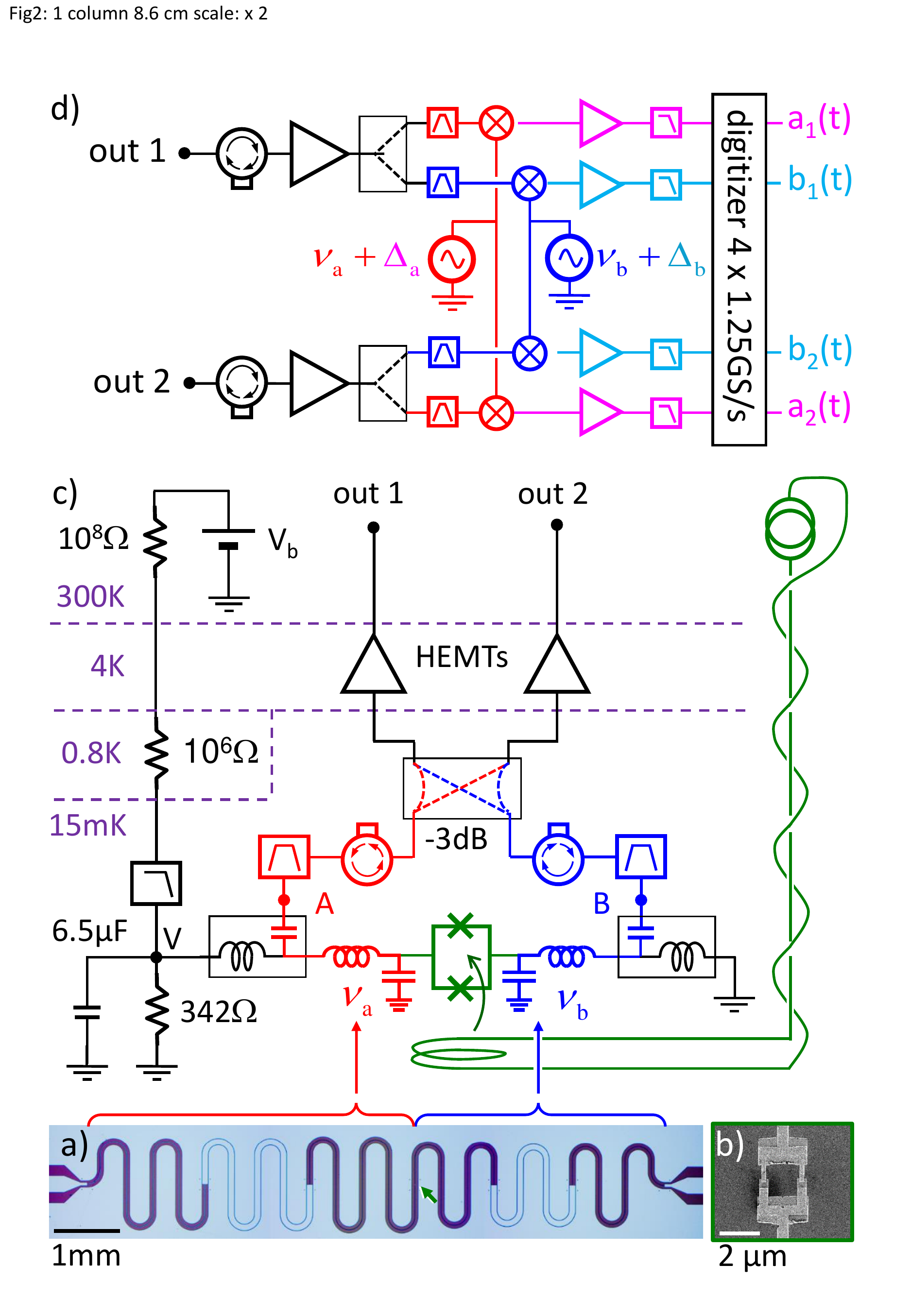}    
\caption{\textbf{Implementation of the experiment sketched in Fig.{\,}\ref{fig1}:} (a) Optical micrograph of the circuit on its silicon chip (see text). The green arrow indicates the position of the SQUID. (b) Scanning electron micrograph of the SQUID used as a tunable Josephson element. (c) Schematic electrical circuit connecting the chip at 15~mK (green SQUID and red and blue LC resonators) to room temperature (see text). (d) Schematic room-temperature setup for heterodyning and digitizing the outgoing fields. The red and blue filters are are $\sim$ 700 MHz wide and centered around $\nu_\mathrm{a}$ and $\nu_\mathrm{b}$, respectively. The microwave signals are then down-converted, amplified and low-pass filtered with a $\sim$ 600 MHz cut-off frequency and then digitized, as shown by the magenta and cyan symbols.}
\label{fig2}
\end{center}
\end{figure}

The implementation of our experiment is shown in Fig.~\ref{fig2}. The Josephson element is implemented as an Al/AlOx/Al superconducting quantum interferometer device [SQUID - see Fig.~\ref{fig2}(b)] whose Josephson energy can be tuned by a magnetic flux applied through its loop; each resonator is made of three cascaded niobium coplanar waveguide segments with different wave impedances [see Fig.~\ref{fig2}(a)]. The sample is anchored at the 15~mK cold stage of a dilution refrigerator and connected [see Fig.~\ref{fig2}(c)] to a low temperature circuitry similar to that described in \cite{rolland2019}. A small current-biased coil applies the tuning magnetic flux. The sample is voltage-biased through two commercial bias-tees (black rectangles) connected to a voltage divider fed by a room-temperature voltage source and heavily filtered at 0.8~K and 15~mK. The two emitted beams are available at the capacitive outputs A and B of the bias-tees. They are then routed through filters and isolators to a -3~dB hybrid coupler that sends both of them to two nominally identical amplification chains 1 and 2, each equipped with a high electron mobility transistor (HEMT) amplifier. This Hanbury Brown-Twiss like setup allows us to reject the uncorrelated noise from the two chains 1 and 2, without affecting the entanglement. Because the $a_\text{out}$ and $b_\text{out}$ field components along 1 and 2 have well separated frequencies, they do not interfere and can be processed by the same wideband HEMT.

The room temperature setup [see Fig.~\ref{fig2}(c)] is designed to measure cross-correlation functions of the signals emitted in two frequency bands around $\nu_\mathrm{a}$ and $\nu_\mathrm{b}$. Each of the signal outputs 1 and 2 is further amplified and split into the two original microwave components (red and blue), which are then filtered and heterodyned to bring them in the dc-600 MHz frequency range (magenta and cyan). The four temporal signals ($a_1$, $b_1$, $a_2$, $b_2$) are finally digitized and post-processed in order to compute power spectral densities (PSD) and  time-domain correlators. The effective acquisition bandwidth of 525~MHz is chosen to be wide enough to capture most of the short-time features of the correlators in inequality (\ref{inequality}).

The two resonances at $\nu_\mathrm{a}=$ 5.092~GHz and $\nu_\mathrm{b}=$ 6.955~GHz as well as their quality factors $Q_\mathrm{a} = 60.8$ and $Q_\mathrm{b}=97.0$, are first obtained \textit{in situ} from shot noise at bias voltage above the superconducting gap \cite{rolland2019}. They lead to similar energy leak rates $\kappa_\mathrm{a}=5.26\times10^8~s^{-1}$ and $\kappa_\mathrm{b}=4.51\times10^8~s^{-1}$. The resonator characteristic impedances yield similar coupling factors $\alpha_\mathrm{a}=0.070$ and $\alpha_\mathrm{b}=0.061$.

\begin{figure}[hb]
\begin{center}
\includegraphics[width = 0.5\textwidth]{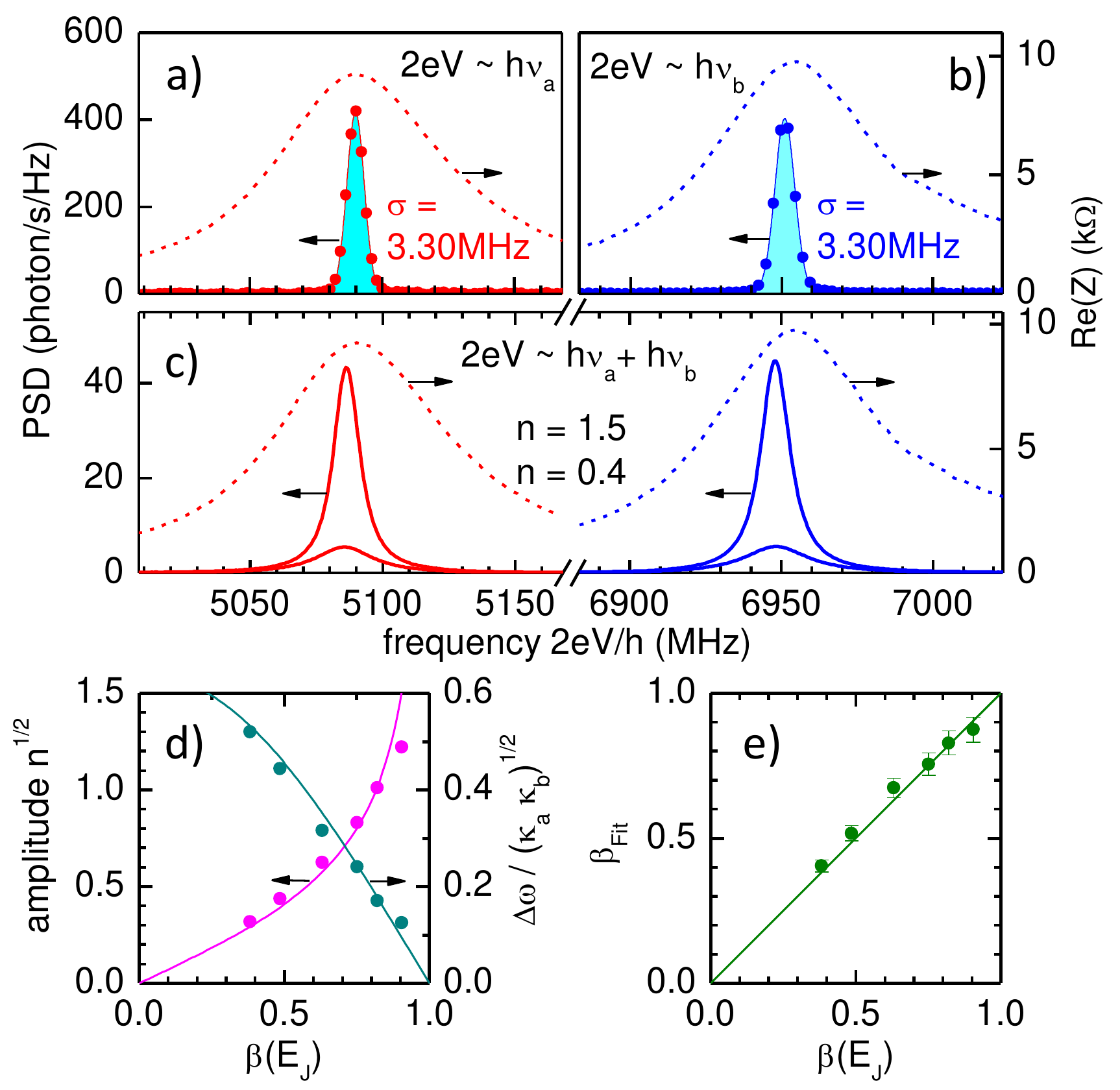}    
\caption{\textbf{Characterization of the emission.}
(a,b) \textit{In situ} measured real part of the impedance across the Josephson element (dashed lines, right axis), and example of measured (dots) and Gaussian-fitted (lines) power spectral densities (left axis) recorded separately around $2\mathrm{e}V=h\nu_\mathrm{a}$ (a) and $2\mathrm{e}V=h\nu_\mathrm{b}$ (b) at small $E_\mathrm{J}$ yielding $n_\mathrm{a}=0.34$ and $n_\mathrm{b}=0.39$, respectively. The emission linewidth results from the voltage-noise induced distribution $p(\nu_J = 2 \mathrm{e}V/h)$ of the Josephson frequency. The corresponding standard deviation $\sigma$ is the same for both lines but drifts slowly back and forth over hours and days.
(c) Same real part of the impedance (dashed lines, right axis) as in (a) , and example of emitted power spectral density (PSD, thick lines) at $2\mathrm{e}V=h\nu_\mathrm{a}+h\nu_\mathrm{b}$ and a driving strengths $\beta=0.67$ and $\beta=0.9$ corresponding to an average photon numbers $n = 0.4$ and $n=1.5$. (d) Measured (symbols) reduced intra-resonator amplitude (red) and  emission width (blue) of the PSDs of Fig.~SM \ref{Supp1}, compared to the corresponding analytical expressions (lines) known for Hamiltonian (\ref{HTMS}). (e)  Dimensionless pumping strength fitted from the same PSD versus calibrated one.
}
\label{fig3}
\end{center}
\end{figure}

\section{Results}

We first calibrate the overall gains of the four acquisition chains ($a_1$, $b_1$, $a_2$, $b_2$) versus frequency $\nu$, and measure the real part of the impedance $\mathrm{Re}[Z(\nu)]$ seen by the Josephson element. The gains [see Fig.~\ref{Supp1}(a) in the Supplemental Material (SM)] display $\sim3$ dB ripples that slightly distort the measured time-domain correlators. The measured $\mathrm{Re}[Z(\nu)]$ shown in Fig.~\ref{fig3}(a,b) displays two Lorentzian lines centered at $\nu_\mathrm{a}$ and $\nu_\mathrm{b}$ and slightly distorted by spurious reflections in the measuring lines.

We then measure the radiation emitted by each resonator in single photon processes corresponding to the AC-Josephson regime, that is when $2\mathrm{e}V \simeq h\nu_\mathrm{a}$ or $h\nu_\mathrm{b}$, with a low-enough Josephson energy to ensure that $\alpha_{\xi} n_{\xi}\ll 1$, thus avoiding stimulated emission effects \cite{westig2017}. An example of the corresponding PSDs is shown in Fig.~\ref{fig3}(a,b). Both emission lines display a Gaussian shape with a standard deviation of $\sigma=3.3$~MHz corresponding to a $h\sigma/2\mathrm{e}=\ 6.8$~nV~rms bias voltage noise. This $\sigma$ value was observed to decrease slowly in time as the experiment was thermalizing, with sudden raises at each liquid helium transfer in the cryostat. Complementary measurements allowed us to attribute this noise mostly to a parasitic LC mode of the bias-tees around 70 kHz, expected to yield $\sigma = \sqrt{2k_B T/C_\mathrm{bias-tee}}$. The lowest observed value $\sigma=2.6$~MHz corresponds to a 20~mK temperature of the bias circuit, in reasonable agreement with the 15~mK temperature of the fridge.

We then set the voltage to the targeted resonance condition $2\mathrm{e}V\simeq h\nu_\mathrm{a}+h\nu_\mathrm{b}$, and observe the simultaneous emission around $\nu_\mathrm{a}$ and $\nu_\mathrm{b}$, for several driving strengths $\beta$ between 0.4 and 1. Figure~\ref{fig3}(c) shows two examples of PSD at $\beta=0.631$ and $\beta=0.905$. The identical shape of the two peaks of the same pair, their common total power $\kappa_\mathrm{a}{n_\mathrm{a}}=\kappa_\mathrm{b}{n_\mathrm{b}}=180$~Mphotons/s and 450~Mphotons/s (yielding an average resonator population $n=\sqrt{n_\mathrm{a} n_\mathrm{b}} = 0.39$ and 1.49), as well as their larger width compared to the single photon case $2\mathrm{e}V \simeq h\nu_{\mathrm{a},\mathrm{b}}$, reflect the emission by photon-pair production.  In Sec. 2 of the SM, we show that all the measured PSDs are well reproduced by the analytical expression computed from the pure two-mode squeezing Hamiltonian (\ref{HTMS}). As shown by Fig. \ref{fig3}(d), we observe a narrowing of the width of the emission spectra with increasing $\beta$, as commonly observed in AC pumped Josephson parametric amplifiers \cite{Roch2012, flurin2012,PhysRevLett.113.110502}. However, as shown by Fig. \ref{fig3}(c) and contrarily to what occurs in these devices, the peak emission frequencies do not shift with increasing pumping strength, provided that the dc voltage $V$ across the junction is kept constant (taking into account the voltage drop across the 342 $\Omega$ output resistance of the biasing circuit shown in Fig. \ref{fig2}). This is due to the absence of Kerr and cross-Kerr non-linearities in the normal-ordered Hamiltonian  (\ref{HRWA}) \cite{SM}. As shown by Fig.~\ref{fig3}(e), fitting the frequency width of the emitted radiation with the two-mode squeezing expressions yield a pumping strength $\beta$ in quantitative agreement with the one deduced from a determination of $E_J^*$ using the AC Josephson effect\cite{SM}.      

The second-order correlators $g^{(2)}_\phi(\tau)$ and $g^{(2)}_\mathrm{ab}(\tau)$ are deduced from the measurements of several correlation functions, as explained in the SM. Figure \ref{fig4}(a) shows the corresponding functions $\left|g^{(2)}_\phi\right|$ and $g^{(2)}_\mathrm{ab,sym}$ for six driving strengths $\beta=$ 0.4, 0.51, 0.67, 0.75, 0.82, 0.9 (and independently measured small detunings $\delta =$~6.7, 2.1, -12.1, 1.8, -11.1, and -11.6~MHz) leading to the average resonator populations $n=$ ~0.10, 0.2, 0.4, 0.7, 1 and 1.5, respectively. These two correlators coincide at $\tau =0$ as expected from their definition. They both present an initial rapid decay over a timescale set by $n$ and the resonators lifetime. The correlator $g^{(2)}_\mathrm{ab,sym}$ converges to 1 while $\left|g^{(2)}_\phi\right|$ subsequently follows a slower Gaussian decay down to zero, as expected from the low-frequency voltage noise already mentioned. The experimental entanglement witness $\left|g^{(2)}_\mathrm{\phi}\right| - g^{(2)}_\mathrm{ab,sym} > 0$ at short time testifies that the two beams at 5 and 7~GHz are entangled. When increasing $\beta$, entanglement was detected up to $n=5$ (data not shown), although full numerical simulations could not be performed at such high occupation numbers (cf. Sec. 5 of SM).

\section{Entanglement analysis}

\begin{figure}[ht]
\begin{center}
\includegraphics[width = 0.5\textwidth]{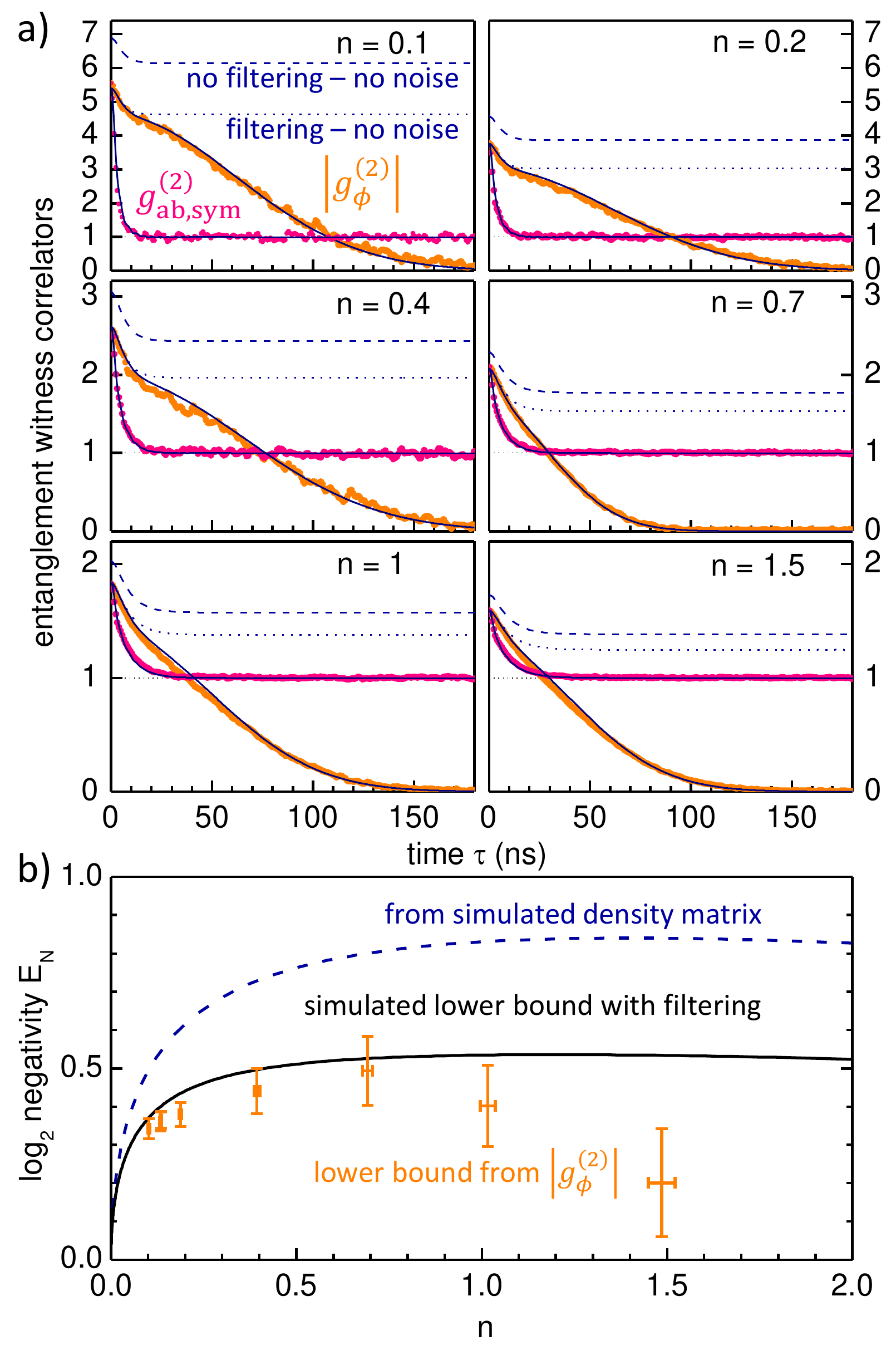}    
\caption{
\textbf{Demonstration of entanglement}.
(a) Measured (dots) and simulated (solid lines) two-times fourth-order in field correlators $g^{(2)}_\mathrm{ab,sym}$ and $g^{(2)}_\phi$ (see text) as a function of delay $\tau$, for different Josephson energies leading to the indicated population $n$. Dashed and dotted lines correspond to simulations without voltage noise and with (dotted) and without (dashed) filtering by the measuring lines.
(b) log2 negativity of the intra-resonator fields as a function of population $n$. Exact value obtained directly from the numerically simulated density matrix of the system (dashed line), and experimental (error bars) and simulated (solid line) lower bound value corresponding to the filtering by the measuring lines (see text).} %(b) The overlapping top lines correspond to simulations with the parameters of the experiment and in absence of any filtering. The blue curve corresponds to the raw negativity of the simulated two-resonator density matrix. The green line is obtained assuming Gaussian beams. The red dashed line is obtained when replacing the experimentally determined couplings and energy decay rates by their geometric averages for the two resonators. The bottom magenta solid line takes into account in the simulation the filtering effect of the experimental setup (see SM). Right axis: Measured (open symbol) and calculated (line) $-3\mathrm{dB}$ emission bandwidth BW  obtained by fitting the power spectral densities (see Figure SM1) and using the simple two-mode squeezing analytical model. Left axis: Computed entanglement rate $\Gamma_E$ in the bandwidth BW (see text) immediately at the output of the resonators, calculated from the simulated negativities of panel (a) The small $r_{a,b}$ and $\kappa_{a,b}$ asymmetry between the two resonators has been neglected. (c) Evaluation of the Gaussian character of the resonator states by simulation. Dots are cuts of the probability distribution $p(U,V)$ of the mixed quadrature $U$ and $V$, at $U=0$ and $V=0$, for a number of photons from $n=0$ to $n=4$ (left to right). Solid lines are Gaussian fit of the quadrature $U$. Dashed lines are Gaussian fits of the other quadrature $V$ imposing the same amplitude.}%, using the simulated $\left|\left< ab\right>\right|$ in formula \ref{log_neg The smallest standard deviation $\sigma U$ drops below that of vacuum, and the Gaussianity is clearly lost between 1 and 2 photons in the resonators. with the smallest standard deviation $\sigma$ (magenta), and 
\label{fig4}
\end{center}
\end{figure}

In order to probe our theoretical modeling of the system dynamics and measurement schemes, we perform numerical quantum simulations of the dynamics of the circuit using the parameters measured in the experiment. The steady-state density matrix $\rho$ of the two resonators is obtained in the Fock state basis using the Lindblad master equation \cite{armour2015} corresponding to Hamiltonian (\ref{HRWA}) and to the relaxation super-operators $\sqrt{\kappa_\mathrm{a}}\hat{a}$ and $\sqrt{\kappa_\mathrm{b}}\hat{b}$. This allows us to check that the field  departs negligibly from Gaussian statistics up to 1 photon in each resonator. Then, using the quantum regression theorem \cite{gardiner2004,breuer2002}, we simulate the dimensionless correlators 
$\left<\hat{b}^\dagger(\tau)\hat{a}^\dagger(0)\hat{a}(0)\hat{b}(\tau)\right>/(n_\mathrm{a} n_\mathrm{b})$ and $\left<\hat{a}^\dagger(0)\hat{b}^\dagger(0)\hat{a}(\tau)\hat{b}(\tau)\right>/(n_\mathrm{a}n_\mathrm{b})$ (see Sec. 5 of SM). Using the standard input-output theory \cite{yurke2004}, one can show that these simulated intra-resonator correlators are actually equal to the expressions (\ref{g2phi}) and (\ref{g2ab}) for the outgoing fields, immediately at the resonator outputs. We thus plot in dashed lines the simulated $\left|g^{(2)}_\phi(\tau)\right|$ functions in the panels of Fig.\ref{fig4}(a), and observe that they are significantly above the measured ones at all times. We improve our simulation of the correlators by including in a simple way (see Secs. 2 and 5 of SM) the previously measured filtering of our amplification chains. We now obtain the dotted lines $\left|g^{(2)}_{\phi,\mathrm{filter}}(\tau)\right|$, which are in good agreement with the experiment at short time, but do not decay to zero at long times. Finally, we also account for the low-frequency voltage fluctuations: we model them as a purely static $\delta$ noise with Gaussian distribution and standard deviation $\sigma$, and compute the weighted average of $g^{(2)}_{\phi,\mathrm{filter}}(\tau)$ over $\delta$. To a very good approximation, this averaging simply multiplies the simulated value by $\exp[-2(\pi\sigma \tau)^2]$. As $\sigma$ was observed to vary slowly in time, we use it as the only fitting parameter to fit each experimental $g_{\phi}^{(2)}(\tau)$. The fits shown in Fig.~\ref{fig4}(a) are now satisfactory at all times and yield $\sigma =$~2.57, 2.61, 2.40, 4.98, 3.10, and 3.64~MHz (in order of increasing $\beta$ or $n$). This good agreement validates our model of voltage noise-induced slow dephasing: on a time scale shorter than $\tau_\mathrm{bias-tee}\sim ~ \sqrt{L_\mathrm{bias-tee}C_\mathrm{bias-tee}} = 2.3~\si{\mu}$s, the pump frequency can be considered as fixed and the junction creates a standard two-mode squeezed state in the resonators.

\section{Log negativity and entanglement witness}

The amount of entanglement between two sub-systems $a$ and $b$ can be quantified by their log-negativity \cite{RevModPhys.81.865}:
\begin{equation}
    E_\mathrm{N} = \log_2(||\rho^{T_b}||_1),
    \label{deflogneg}
\end{equation}
where $\rho^{T_b}$ is the partial transpose of the density operator $\rho$ of the total system with respect to subsystem $b$, and $||.||_1$ is the trace norm. It is an upper bound on the number of entangled bits (or e-bits) that can be distilled from the system. This quantity is directly available from our numerical simulations of the intra-resonator fields. It is plotted as a function of $n$ in Fig. \ref{fig4}(b): $E_\mathrm{N}$ increases rapidly to a very flat maximum of about $~0.8$\,e-bits between $n=1$ and 2 photons. 

In the case of a two-mode squeezed Gaussian state of $a$ and $b$, $E_\mathrm{N}$ can be linked to observable quantities through:
\begin{equation}
    E_\mathrm{N} = \log_2{\left[1+n_\mathrm{a}+n_\mathrm{b}-\sqrt{(n_\mathrm{a}-n_\mathrm{b})^2+4\left|\left<ab\right>\right|^2}\right].}
\label{log_neg}
\end{equation}

 We first checked that in our simulations using the full Hamiltonian Eq. \ref{HRWA}, the values of $E_N$ computed using Eq. \ref{log_neg} agreed with the one computed using Eq. \ref{deflogneg} with minute deviations. This allows us to deduce $E_\mathrm{N}$ from the $g^{(2)}_\phi(\tau)$ function measured on the outgoing fields. For a noiseless voltage bias, the correlator in Eq. (\ref{log_neg}) is given by $\left|\left<\hat{a}_\mathrm{out}\hat{b}_\mathrm{out}\right>\right|^2 = \kappa_\mathrm{a} \kappa_\mathrm{b}\left|\left<\hat{a}\hat{b}\right>\right|^2=n_\mathrm{a} n_\mathrm{b} g^{(2)}_\phi(\tau\rightarrow\infty)$. Although the measured $g^{(2)}_\phi(t\rightarrow\infty)$ vanishes due to quasi-static noise, we can retrieve its corresponding noiseless values by dividing it by the Gaussian envelopes fitted above (see SM sections 3). The corresponding apparent log-negativities, shown by the orange error bars in Fig. \ref{fig4}(b), are 40\% below the theoretical estimate, shown by the blue dashed line in Fig. \ref{fig4}(b). This discrepancy is explained by the fact that they are only lower bounds of $E_\mathrm{N}$, due to the filtering by the lines. This is confirmed by simulations of $\left<\hat{a}_\mathrm{out}\hat{b}_\mathrm{out}\right>$ taking into account the measured filtering, which yield to a $E_\mathrm{N}(n)$ curve, shown by the black solid line in Fig. \ref{fig4}(b), in better overall agreement with the measurements. However, the two points at 1 and 1.5 photons are markedly below our prediction, which we attribute to a stronger coupling of the Josephson junction to its low frequency environment. Indeed, when the detuning $\delta$ is positive (negative), the dc differential conductance of the junction happens to be negative (positive), which amplifies (cools down) the low frequency noise of its environment. At positive detuning  (and even zero detuning in presence of large voltage fluctuations), the amplifying effect makes the system unstable and hysteretic, all the more easily that the drive $\beta$ (and thus $n$) is large. To avoid this instability, we actually applied slightly negative detunings to measure the points at $n=1$ and $n=1.5$. The junction then cools down the low frequency environment \cite{PhysRevB.94.235420}, converting its energy to the much higher frequencies $\nu_{\mathrm{A},\mathrm{B}}$, and degrading the detected entanglement. A systematic and quantitative investigation of this effect goes beyond the scope of the present work, and is left for further investigations.

Although a quantitative account could be reached for the entanglement witness probed, the absence of a stable phase reference for the two-mode squeezing does not allow us measure sufficiently rapidly the entanglement entropy of each pair of outgoing modes $(\nu, \nu’=2\mathrm{e}V/h-\nu)$ \cite{Deng_2016}.  It is nevertheless interesting to estimate the instantaneous flux of entanglement entropy,   during  microsecond long stability periods. We use for this purpose  the  squeezing Hamiltonian model (\ref{HTMS}) that yields emission [see Fig.~\ref{fig3}(d,e)]  and entanglement properties (see Fig.~\ref{fig4}) in agreement with  the experimental data.  As described in the Supplemental Material \cite{SM}, we calculate  the entanglement spectral density $E_\mathrm{N}(\nu)$ defined  in \cite{Deng_2016} or the related density of entropy of formation $E_\mathrm{F}(\nu)$ \cite{PhysRevA.72.032334}, and integrate it over the emission bandwidth $\Delta \nu$.  This procedure yields a total entanglement ﬂux $\Gamma_\mathrm{N}(\nu)$ and $\Gamma_\mathrm{F}(\nu)$ of about 115 and  103 Mebit/s, respectively. However, we stress here that this appealing figure of merit compared to the one achieved with parametric amplifiers or converters \cite{flurin2012,PhysRevLett.124.140503,Fedorov_2018} is only an estimate and is not exploitable by the quantum information protocols presently known, due to diffusion of the squeezing angle. A possible way to directly detect the entanglement between the $(\nu, \nu’)$ outgoing modes would be to incorporate an additional Josephson junction sharing the same DC bias and to use its Josephson radiation as a phase reference.

\section{Discussion}

We now discuss the advantages and limitations of our circuit and its potential developments. We first stress that our dc-biased circuit does not suffer from the Kerr non-linearities observed in ac-pumped Josephson parametric amplifiers and converters \cite{Roch2012, flurin2012,PhysRevLett.113.110502}. The non-linearity of Hamiltonian (\ref{HRWA}) is of a different nature and is always easy to repel at higher photon numbers: In our experiment, the Gaussian character of the emitted light was ensured up to only 1-1.5 photons in the resonators ($\alpha_{\xi} \langle \hat{n}_{\xi} \rangle \ll 1$), but lowering by a factor 10 the impedance of the two-modes (and consequently of the couplings $\alpha_{\mathrm{a},\mathrm{b}}$),  while increasing the Josephson energy $E_\mathrm{J}$ by the same factor, one could make our entangled microwave source 10 times brighter. Conversely, coupling a dc biased Josephson junction to several high impedance resonators such as the one described in \cite{rolland2019} would generate highly non-Gaussian entangled beams of photons \cite{PhysRevX.10.011011}, and combining a high and a low impedance mode has been predicted to stabilize a Fock state of the high impedance mode with a mere DC bias using the two photons processes exploited in this paper \cite{souquet2016}. Using Josephson junctions with higher superconducting gaps, such as NbN \cite{grimm2019}, should allow to observe the entanglement evidenced here up to THz frequencies. 
Last, transforming our circuit into a useful source of continuous entangled beams, requires maintaining the stability of the squeezed two-mode quadrature over long enough times. This implies a signiﬁcant reduction of the noise on the bias-voltage  $V$, which could be done on-chip using a Shapiro step of an additional Josephson junction.

In a broader picture, it is worth noting that the entanglement of the outgoing modes originates from successive two-photon emission processes associated to Cooper pair tunneling between superconducting condensates with a controlled superconducting phase diﬀerence.  The properties of the radiation emitted by an out-of equilibrium quantum conductor arise from charge quantization and from the quantum correlations of its electronic reservoirs. The present work is thus a prime example of the emerging field of mesoscopic quantum electrodynamics of coherent conductors where numerous interesting phenomena were already recently predicted and demonstrated for producing e.g. sub-Poissonian photon sources \cite{beenakker2001,beenakker2004,fulga2010,fulga2010, hassler2015, grimm2019,  rolland2019, PhysRevB.100.054515}, novel types of lasers \cite{cassidy2017,PhysRevLett.107.073901,PhysRevB.87.094511, PhysRevB.89.104502}, near-quantum limited amplifiers \cite{jebari2018,PhysRevApplied.11.034035}, squeezed radiation \cite{ forgues2015,grimsmo2016,dmytruk2016,mora2017}, or new types of Q-Bits \cite{ esteve2018}, or two-photon losses \cite{Cottet2019}.

%\section{Conclusion}
As a conclusion, by measuring an entanglement witness, we were able to show that from a purely classical voltage source, a non-linear active element was capable to emit into its environment a flux of entangled microwave beams at different frequencies. We extracted a lower bound on the value of the logarithmic negativity and showed that entanglement is conserved up to 5 photons present in each resonators, corresponding to a flux of 2.5 billions photon pairs per second.

We gratefully acknowledge stimulating discussions with C. Padurariu, S. Wölk, A. Aspect, E. Flurin, F. Grosshans and N. Treps, and precious technical help from P. Jacques.  This work received funding from the European Research
Council under the European Unions Program for Research and Innovation (Horizon
2020)/ERC Grant Agreement No. [639039].
We gratefully acknowledge partial support from
LabEx PALM (ANR-10-LABX-0039-PALM),
ANR contract GEARED and SIM-CIRCUIT,
from the ANR-DFG Grant JosephCharli, and
from the ERC through the NSECPROBE grant,
from IQST and the German Science Foundation (DFG) through AN336/11-1. S.D. acknowledges financial support by the Carl Zeiss Foundation and the German Academic Exchange Service (DAAD).  

\bibliography{biblio}

\end{document}